\documentclass[conference]{IEEEtran}
\usepackage[mathscr]{eucal}
\usepackage[cmex10]{amsmath}
\usepackage{epsfig,epsf,psfrag}
\usepackage{amssymb,amsmath,amsthm,amsfonts,latexsym}
\usepackage{amsmath,graphicx,bm,xcolor,url,overpic}
\usepackage{fixltx2e}%ordering of single and double column floats
\usepackage{array}%array and tabular environments
\usepackage{verbatim}
\usepackage{bm}
\usepackage{algorithmic}
\usepackage{algorithm}
\usepackage{verbatim}
\usepackage{textcomp}
\usepackage{mathrsfs}
\usepackage{epstopdf}

\newcommand{\openone}{\leavevmode\hbox{\small1\normalsize\kern-.33em1}}

\catcode`~=11 \def\UrlSpecials{\do\~{\kern -.15em\lower .7ex\hbox{~}\kern .04em}} \catcode`~=13 

\allowdisplaybreaks[4]

\newcommand{\nn}{\nonumber}

\newcommand{\calA}{\mathcal{A}}

\newcommand{\calF}{\mathcal{F}}

\newcommand{\calP}{\mathcal{P}}

\newcommand{\calS}{\mathcal{S}}

\newcommand{\calX}{\mathcal{X}}
\newcommand{\calY}{\mathcal{Y}}

\newcommand{\ba}{\mathbf{a}}

\newcommand{\bS}{\mathbf{S}}

\newcommand{\bw}{\mathbf{w}}

\newcommand{\bx}{\mathbf{x}}
\newcommand{\bX}{\mathbf{X}}

\newcommand{\rma}{\mathrm{a}}

\newcommand{\rmb}{\mathrm{b}}

\newcommand{\rmd}{\mathrm{d}}

\newcommand{\rme}{\mathrm{e}}

\newcommand{\rmP}{\mathrm{P}}

\newcommand{\bbN}{\mathbb{N}}

\newcommand{\bbR}{\mathbb{R}}

\DeclareMathAlphabet{\mathbsf}{OT1}{cmss}{bx}{n}
\DeclareMathAlphabet{\mathssf}{OT1}{cmss}{m}{sl}% slanted sans serif

\DeclareSymbolFont{bsfletters}{OT1}{cmss}{bx}{n}  
\DeclareSymbolFont{ssfletters}{OT1}{cmss}{m}{n}
\DeclareMathSymbol{\bsfGamma}{0}{bsfletters}{'000}
\DeclareMathSymbol{\ssfGamma}{0}{ssfletters}{'000}
\DeclareMathSymbol{\bsfDelta}{0}{bsfletters}{'001}
\DeclareMathSymbol{\ssfDelta}{0}{ssfletters}{'001}
\DeclareMathSymbol{\bsfTheta}{0}{bsfletters}{'002}
\DeclareMathSymbol{\ssfTheta}{0}{ssfletters}{'002}
\DeclareMathSymbol{\bsfLambda}{0}{bsfletters}{'003}
\DeclareMathSymbol{\ssfLambda}{0}{ssfletters}{'003}
\DeclareMathSymbol{\bsfXi}{0}{bsfletters}{'004}
\DeclareMathSymbol{\ssfXi}{0}{ssfletters}{'004}
\DeclareMathSymbol{\bsfPi}{0}{bsfletters}{'005}
\DeclareMathSymbol{\ssfPi}{0}{ssfletters}{'005}
\DeclareMathSymbol{\bsfSigma}{0}{bsfletters}{'006}
\DeclareMathSymbol{\ssfSigma}{0}{ssfletters}{'006}
\DeclareMathSymbol{\bsfUpsilon}{0}{bsfletters}{'007}
\DeclareMathSymbol{\ssfUpsilon}{0}{ssfletters}{'007}
\DeclareMathSymbol{\bsfPhi}{0}{bsfletters}{'010}
\DeclareMathSymbol{\ssfPhi}{0}{ssfletters}{'010}
\DeclareMathSymbol{\bsfPsi}{0}{bsfletters}{'011}
\DeclareMathSymbol{\ssfPsi}{0}{ssfletters}{'011}
\DeclareMathSymbol{\bsfOmega}{0}{bsfletters}{'012}
\DeclareMathSymbol{\ssfOmega}{0}{ssfletters}{'012}

\newcommand{\hats}{\hat{s}}
\newcommand{\hatS}{\hat{S}}

\newcommand{\hatt}{\hat{t}}

\newcommand{\hatw}{\hat{w}}

\newtheorem{theorem}{Theorem}

\newtheorem{definition}{Definition}

\usepackage[utf8]{inputenc} % allow utf-8 input
\usepackage[T1]{fontenc}    % use 8-bit T1 fonts
\usepackage{url,bbm}            % simple URL typesetting
\usepackage{booktabs}       % professional-quality tables
\usepackage{amsfonts}       % blackboard math symbols
\usepackage{nicefrac}       % compact symbols for 1/2, etc.
\usepackage{microtype}      % microtypography

\usepackage{cite}
\usepackage{subfigure,transparent,color,graphicx} %used to insert pictures

\graphicspath{{./figures2/}}
\allowdisplaybreaks[2]
\newcommand{\bbo}{\mathbbm{1}}
\IEEEoverridecommandlockouts
\begin{document}

\title{Achievable Resolution Limits for the Noisy Adaptive 20 Questions Problem}
\author{  
 \IEEEauthorblockN{Lin Zhou}
  \IEEEauthorblockA{School of Cyber Science and Technology\\
Beihang University\\
    Email: \url{lzhou@buaa.edu.cn} }
  \and
  \IEEEauthorblockN{Alfred Hero}
  \IEEEauthorblockA{Department of EECS\\
       University of Michigan \\
   Email: \url{hero@eecs.umich.edu}}
   \thanks{This work was supported in part by the National Key Research and Development Program of China under Grant 2020YFB1804800 and in part by ARO grant W911NF-15-1-0479.}
}
\maketitle

\begin{abstract}
We study the achievable performance of adaptive query procedures for the noisy 20 questions problem with measurement-dependent noise over a unit cube of finite dimension. The performance criterion that we consider is the minimal resolution, defined as the $L_\infty$ norm between the estimated and the true values of the random location vector of a target, given a finite number of queries constrained by an excess-resolution probability. Specifically, we derive the achievable resolution of an adaptive query procedure based on the variable length feedback code by Polyanskiy \emph{et al.} (TIT 2011). Furthermore, we verify our theoretical results with numerical simulations and compare the performance of our considered adaptive query procedure with that of certain state-of-the-art algorithms, such as the sorted posterior matching algorithm by Chiu and Javadi (ITW 2016). In particular, we demonstrate that the termination strategy adopted in our adaptive query procedure can significantly enhance the asymptotic performance of adaptive query procedures, especially at moderate to large excess-resolution probability constraints.
\end{abstract}

\section{Introduction}
The noisy 20 questions problem (cf. \cite{renyi1961problem,burnashev1974,ulam1991adventures,pelc2002searching,jedynak2012twenty,chung2018unequal,lalitha2018improved}) arises when one aims to accurately estimate an arbitrarily distributed random variable $S$ by successively querying an oracle and using its noisy responses to form an estimate $\hatS$. A central goal in this problem is to find optimal query strategies that yield a good estimate $\hatS$ for the unknown target random variable $S$.

Depending on the query design strategy adopted, the 20 questions problem can either be adaptive or non-adaptive. In adaptive query procedures, the design of a subsequent query depends on all previous queries and noisy responses to these queries from the oracle. In non-adaptive query procedures, all the queries are designed independently in advance. For example, the bisection policy~\cite[Section 4.1]{jedynak2012twenty} is an adaptive query procedure and the dyadic policy~\cite[Section 4.2]{jedynak2012twenty} is a non-adaptive query procedure. Depending on whether or not the structure of the noise depends on the queries, the noisy 20 questions problem is classified into two categories: querying with measurement-independent noise (e.g.,~\cite{jedynak2012twenty,chung2018unequal}); and querying with measurement-dependent noise (e.g.,~\cite{kaspi2018searching,lalitha2018improved}). As argued in \cite{kaspi2018searching}, measurement-dependent noise better models practical applications. For example, for target localization in a sensor network, the noisy response to each query can depend on the size of the query region due to possible presence of clutter. Another example is in human query systems where personal biases about the state may affect the response.

In earlier works on the noisy 20 questions problem, e.g.,~\cite{jedynak2012twenty,tsiligkaridis2014collaborative,tsiligkaridis2015decentralized}, the queries were designed to minimize the entropy of the posterior distribution of an one-dimensional target variable $S$. As pointed out in later works, e.g., \cite{chung2018unequal,chiu2016sequential,kaspi2018searching,lalitha2018improved}, other accuracy measures, such as the estimation resolution and the quadratic loss are often better criteria for localization, where the resolution is defined as the absolute difference between $S$ and its estimate $\hatS$ and the quadratic loss is $(\hatS-S)^2$.

In our previous work~\cite{zhouisit2020}, we derived non-asymptotic and asymptotic bounds on the minimal achievable resolution of an optimal non-adaptive query procedure (see also \cite[Theorems 1-3]{zhou2019twentyq}). One may then wonder whether adaptive query procedures have strict benefit over non-adaptive query procedures, either non-asymptotically or asymptotically. In this work, we provide an affirmative answer to this question by deriving the achievable resolution of an adaptive query procedure based on the variable length feedback code in \cite[Definition 1]{polyanskiy2011feedback} for a multidimensional target. Furthermore, when specialized to a one-dimensional target and a measurement-dependent binary symmetric channel, we compare the non-asymptotic and asymptotic performances of our adaptive query procedure with those of certain state-of-the-art algorithms such as the sorted posterior matching algorithm in \cite{chiu2016sequential}.

\section{Problem Formulation}
\subsection*{Notation}
Random variables and their realizations are denoted by upper case variables (e.g.,  $X$) and lower case variables (e.g.,  $x$), respectively. All sets are denoted in calligraphic font (e.g.,  $\mathcal{X}$). Let $X^n:=(X_1,\ldots,X_n)$ be a random vector of length $n$. We use $\Phi^{-1}(\cdot)$ to denote the inverse of the cumulative distribution function (cdf) of the standard Gaussian. We use $\bbR$, $\bbR_+$ and $\bbN$ to denote the sets of real numbers, positive real numbers and integers respectively. For any real number $p\in(0,1)$, we use $\mathrm{Bern}_p(\cdot)$ to denote the Bernoulli distribution with parameter $p$, i.e., if $X\sim\mathrm{Bern}_p(\cdot)$, then $\Pr\{X=1\}=p$. Given any two integers $(m,n)\in\bbN^2$, we use $[m:n]$ to denote the set of integers $\{m,m+1,\ldots,n\}$ and use $[m]$ to denote $[1:m]$. Given any $(m,n)\in\bbN^2$, for any $m$ by $n$ matrix $\ba=\{a_{i,j}\}_{i\in[m],j\in[n]}$, the infinity norm is defined as $\|\ba\|_{\infty}:=\max_{i\in[m],j\in[n]}|a_{i,j}|$. The set of all probability distributions on a finite set $\calX$ is denoted as $\calP(\calX)$ and the set of all conditional probability distributions from $\calX$ to $\calY$ is denoted as $\calP(\calY|\calX)$. Furthermore, we use $\calF(\calS)$ to denote the set of all probability density functions on a set $\calS$. All logarithms are base $e$. Finally, we use $\bbo()$ to denote the indicator function.

\subsection{Noisy 20 Questions Problem}
Consider an arbitrary integer $d\in\bbN$. Let $\bS=(S_1,\ldots,S_d)$ be a continuous random vector defined on the unit cube of dimensional $d$ (i.e., $[0,1]^d$) with arbitrary probability density function (pdf) $f_{\bS}$. Note that any searching problem over a bounded $d$-dimensional rectangular region with different lengths in each dimension is equivalent to a searching problem over the unit cube of dimension $d$ with normalization in each dimension. 

In the estimation problem formulated under the adaptive query framework of noisy 20 questions, a player aims to accurately estimate the target random variable $\bS$ by sequentially posing queries to an oracle knowing $\bS$. At each time point $i\in\bbN$, the player poses a query asking whether the target lies in a Lebesgue measurable query set $\calA_i\subseteq[0,1]$, where the design of $\calA_i$ depends on previous queries $(\calA_1,\ldots,\calA_{i-1})$ and the noisy responses to these queries. After receiving the query $\calA_i$, the oracle finds the binary answer $X_i=\bbo(\bS\in\calA_i)$ and passes the answer through a measurement-dependent channel yielding the noisy response $Y_i$. Given the accumulated noisy responses, the player decides whether to stop querying and making an estimate $\hat{\bS}=(\hatS_1,\ldots,\hatS_d)$ with a decoding function $g_i:\calY^i\to[0,1]^d$ or to pose further queries to obtain further information. Throughout the paper, we assume that the alphabet $\calY$ for the noisy response is finite. 

Note that in adaptive querying, the player needs to choose a stopping criterion, which may be random, determining the number of queries to make. A formal definition of an adaptive query procedure is provided in Definition \ref{def:adaptive:procedure}.

\vspace{-0.2em}
\subsection{The Measurement-Dependent Channel}
\label{sec:mdc}
We then briefly describe the measurement-dependent channel~\cite{kaspi2018searching,chiu2016sequential}, also known as a channel with state~\cite[Chapter 7]{el2011network}. Given a query $\calA\subseteq[0,1]^d$, the channel from the oracle to the player is denoted by $P_{Y|X}^\calA\in\calP(\calY|\{0,1\})$. Define the size $|\calA|$ of $\calA$ as its Lebesgue measure, i.e., $|\calA|=\int_{t\in\calA}\rmd t$. We assume that the measurement-dependent channel $P_{Y|X}^{\calA}$ depends on the query $\calA$ only through its size, i.e., $P_{Y|X}^{\calA}$ is equivalent to a channel with state $P_{Y|X}^q$ where the state $q$ is a function of $|\calA|$. 

We consider the case where the state of the channel is $q=f(|\calA|)$ where the function $f:[0,1]\to\bbR_+$ is a bounded Lipschitz continuous function with parameter $K$, i.e., $|f(q_1)-f(q_2)|\leq K|q_1-q_2|$ and $\max_{q\in[0,1]}f(q)<\infty$.  A simple choice of $f$ is the identity function, i.e., $f(|\calA|)=|\calA|$ and for this case $K=1$. Furthermore, it is interesting to note that when $f(|\calA|)$ equals a constant for any query $\calA\subseteq[0,1]^d$, the above measurement-dependent model actually reduces to a \emph{measurement-independent} channel where the noisy channel remains the same regardless of the query posed to the oracle.

For any $q\in[0,1]$, any $\xi\in(0,\min(q,1-q))$, we assume that the measurement-dependent channel is continuous in the sense that there exists a constant $c(q)$ depending on $q$ only such that
\begin{align}
\max\left\{\left\|\log\frac{P_{Y|X}^q}{P_{Y|X}^{{q+\xi}}}\right\|_{\infty},\left\|\log\frac{P_{Y|X}^q}{P_{Y|X}^{q-\xi}}\right\|_{\infty}\right\}\leq c(q)\xi\label{assump:continuouschannel}.
\end{align}
Here we recall one example of the measurement-dependent channel that data back to \cite{chiu2016sequential}.
\begin{definition}
\label{def:mdBSC}
Given any $\calA\subseteq[0,1]$, a channel $P_{Y|X}^{\calA}$ is said to be a measurement-dependent Binary Symmetric Channel (BSC) with parameter $\nu\in(0,1]$ if $\calX=\calY=\{0,1\}$ and 
\begin{align}
P_{Y|X}^{\calA}(y|x)=(\nu f(|\calA|))^{\bbo(y\neq x)}(1-\nu f(|\calA|))^{\bbo(y=x)},
\end{align}
for any $(x,y)\in\{0,1\}^2$.
\end{definition}
Note that the output of a measurement-dependent BSC with parameter $\nu$ is the same as the input with probability $1-\nu f(|\calA|)$ and flipped with probability $\nu f(|\calA|)$. It can be verified that the constraint in \eqref{assump:continuouschannel} is satisfied for the measurement-dependent BSC. For this case, a valid choice of the function $f(\cdot)$ should satisfy $\nu f(|\calA|)\leq 1$ for any $\calA\subseteq[0,1]$. In particular, we only consider Lipschitz continuous function $f(\cdot)$ so that $\nu f(|\calA|)\leq \frac{1}{2}$ for any $\calA\subseteq[0,1]$. This is because having a crossover probability greater than $\frac{1}{2}$ is impractical for a BSC.

\vspace{-0.2em}
\subsection{Adaptive Query Procedures}
An adaptive query procedure with resolution $\delta$ and excess-resolution constraint $\varepsilon$ is defined as follows.
\begin{definition}
\label{def:adaptive:procedure}
Given any $(l,d,\delta,\varepsilon)\in\bbR_+\times\bbN\times\bbR_+\times[0,1]$, an $(l,d,\delta,\varepsilon)$-adaptive query procedure for the noisy 20 questions problem consists of
\begin{itemize}
\item a sequence of adaptive queries where for each $i\in\bbN$, the design of query $\calA_i\subseteq[0,1]^d$ depends on all previous queries $\{\calA_j\}_{j\in[i-1]}$ and the noisy responses $Y^{i-1}$ from the oracle
\item a sequence of decoding functions $g_i:\calY^i\to[0,1]^d$ for $i\in\bbN$ 
\item a random stopping time $\tau$ depending on noisy responses $\{Y_i\}_{i\in\bbN}$ such that under any pdf $f_{\bS}$ of the target random variable $\bS$, the average number of queries satisfies
\begin{align}
\mathbb{E}[\tau]\leq l,
\end{align}
\end{itemize}
such that the excess-resolution probability satisfies
\begin{align}
\nn&\rmP_{\rme,\rma}(l,d,\delta)\\*
&:=\sup_{f_{\bS}\in\calF([0,1]^d)}\Pr\{\exists~i\in[d]:~|\hatS_i-S_i|>\delta\}\leq \varepsilon\label{def:excessresolution},
\end{align}
where $\hatS_i$ is the estimate of $i$-th element of the target $\bS$ using the decoder $g$ at time $\tau$, i.e., $g(Y^\tau)=(\hatS_1,\ldots,\hatS_d)$.
\end{definition}
Examples of adaptive query procedures include Algorithm \ref{procedure:adapt} and certain state-of-the-art algorithm such as the sorted posterior matching algorithm in \cite{chiu2016sequential} that builds on the result of Burnashev and Zigangirov in \cite{burnashev1974}.

Given any $(l,d,\varepsilon)\in\bbR_+\times\bbN\times[0,1)$, we can define the fundamental resolution limit for adaptive querying as follows: 
\begin{align}
\delta_\rma^*(l,d,\varepsilon)
\nn&:=\inf\{\delta\in\bbR_+:~\exists~\mathrm{an}~(l,d,\delta,\varepsilon)\mathrm{-adaptive}\\*
&\qquad\qquad\qquad\qquad\qquad\mathrm{query~procedure}\}\label{def:delta*:adaptive},
\end{align}
with analogous definition of mean sample complexity
\begin{align}
l^*(d,\delta,\varepsilon)
\nn&:=\inf\{l\in\bbR_+:~\exists~\mathrm{an}~(l,d,\delta,\varepsilon)\mathrm{-adaptive}\\*
&\qquad\qquad\qquad\qquad\qquad\mathrm{query~procedure}\}.
\end{align}
Note that $l^*(d,\delta,\varepsilon)$ is simply a function of $\delta_\rma^*(l,d,\varepsilon)$ and thus it suffices to study $\delta_\rma^*(l,d,\varepsilon)$. Furthermore, recall that the fundamental limit $\delta^*(n,d,\varepsilon)$ for non-adaptive querying was defined in \cite[Eq. (7)]{zhou2019twentyq} where $n$ is the predetermined number of queries.

\section{Main Results}
The proof of theoretical results are omitted due to space limitation and similarity to the proofs of the simple case of $f(|\calA|)=|\calA|$ in \cite{zhou2019twentyq}. The main purpose of this paper is to compare the performance of the adaptive query procedure in Algorithm \ref{procedure:adapt2} and the sorted posterior matching algorithm in \cite{chiu2016sequential}, which was left as future work in \cite{zhou2019twentyq}.

\subsection{Preliminaries}
Given any $(p,q)\in[0,1]^2$, let $P_Y^{p,q}$ be the marginal distribution on $\calY$ induced by the Bernoulli distribution $P_X=\mathrm{Bern}(p)$ and the measurement-dependent channel $P_{Y|X}^{q}$. Furthermore, define the following information density
\begin{align}
\imath_{p,q}(x;y)&:=\log\frac{P_{Y|X}^q(y|x)}{P_Y^{p,q}(y)},~\forall~(x,y)\in\calX\times\calY.
\end{align}
Correspondingly, given any $n\in\bbN$, for any $(x^n,y^n)\in\calX^n\times\calY^n$, we define
\begin{align}
\imath_{p,q}(x^n;y^n)
&:=\sum_{i\in[n]}\imath_{p,q}(x_i;y_i)\label{def:ixnyn}
\end{align}
as the mutual information density between $x^n$ and $y^n$.

Let $\bX^\infty$ be a collection of $M^d$ random binary vectors $\{X^\infty(i_1,\ldots,i_d)\}_{(i_1,\ldots,i_d)\in[M]^d}$, each with infinite length and let $\bx^\infty$ denote a realization of $\bX^\infty$. Furthermore, let $Y^\infty$ be another random vector with infinite length where each element takes values in $\calY$ and let $y^\infty$ be a realization of $Y^\infty$. For any vector $\bw=(w_1,\ldots,w_d)\in[0,1]^d$ and any integer $n\in\bbN$, given any sequence of queries $\calA^n=(\calA_1,\ldots,\calA_n)\in[0,1]^d$, define the following joint distribution of $(\bX^n,Y^n)$
\begin{align}
P_{\bX^tY^t}^{\calA^t,\bw}(\bx^t,y^t)
\nn&=\prod_{t\in[n]}\Big(\prod_{(i_1,\ldots,i_d)\in[M]^d}\mathrm{Bern}_p(x_t(i_1,\ldots,i_d))\Big)\\*
&\qquad\qquad\times P_{Y|X}^{\calA_t}(y_t|x_t(\bw))\label{def:pxyan}.
\end{align}
We can define $P_{\bX^\infty,Y^\infty}^{\calA^n,\bw}$ as a generalization of $P_{\bX^n,Y^n}^{\calA^n,\bw}$ with $n$ replaced by $\infty$. Since the channel is memoryless, such a generalization is reasonable.

Given any $(d,M)\in\bbN^2$, define a function $\Gamma:[M]^d\to [M^d]$ as follows: for any $(i_1,\ldots,i_d)\in[M]^d$,
\begin{align}
\Gamma(i_1,\ldots,i_d)=1+\sum_{j\in[d]}(i_j-1)M^{d-j}\label{def:Gamma}.
\end{align}
Note that the function $\Gamma(\cdot)$ is invertible. We denote $\Gamma^{-1}:[M^d]\to [M]^d$ the inverse function. Furthermore, given any $\lambda\in\bbR_+$ and any $m\in[M^d]$, define the stopping time
\begin{align}
\tau_m(\bx^\infty,y^\infty)&:=\inf\{n\in\bbN:~\imath_q(x^n(\Gamma^{-1}(m));y^n)\geq \lambda\}\label{def:taum}.
\end{align}

\subsection{Main Results and Discussions}
\begin{algorithm}[bt]
\caption{Adaptive query procedure}
\label{procedure:adapt}
\begin{algorithmic}
\REQUIRE Three parameters $(M,q,\lambda)\in\bbN\times(0,1)\times\bbR_+$
\ENSURE An estimate $(\hats_1,\ldots,\hats_d)\in[0,1]^d$ of a $d$-dimensional target variable $(s_1,\ldots,s_d)\in[0,1]^d$\\
\hrulefill
\STATE Partition the unit cube of dimension $d$ (i.e., $[0,1]^d$) into $M^d$ equal-sized disjoint regions $\{\calS_{i_1,\ldots,i_d}\}_{(i_1,\ldots,i_d)\in[M]^d}$.
\STATE $t \leftarrow 1$
\WHILE {$t>0$}
\STATE Generate $M^d$ binary random variables $\{x_t(i_1,\ldots,i_d)\}_{(i_1,\ldots,i_d)\in[M]^d}$ independently from a Bernoulli distribution with parameter $q$.
\STATE Form the $t$-th query as
\begin{align*}
\calA_t:=\bigcup_{(i_1,\ldots,i_d)\in[M]^d:x_t(i_1,\ldots,i_d)=1}\calS_{i_1,\ldots,i_d}.
\end{align*}
\STATE Obtain the noisy response $y_t$ from the oracle to the query $\calA_t$.
\STATE Calculate accumulated mutual information densities $\imath_{q,f(q)}(x^t(i_1,\ldots,i_d);y^t)$ for all $(i_1,\ldots,i_d)\in[M]^d$.
\IF {$\max_{(i_1,\ldots,i_d)\in[M]^d}\imath_{q,f(q)}(x^t(i_1,\ldots,i_d);y^t)\geq \lambda$}
\STATE $\tau\leftarrow t$.
\STATE $t\leftarrow 0$.
\ELSE
\STATE $t\leftarrow t+1$.
\ENDIF
\ENDWHILE
\STATE Generate estimates $(\hats_1,\ldots,\hats_d)$ as
\begin{align*}
\hats_i=\frac{2\hatw_i-1}{2M},
\end{align*}
where $\hat{\bw}=(\hatw_1,\ldots,\hatw_d)$ is obtained as follows:
\begin{align*}
\hat{\bw}&=\Gamma^{-1}(\hatt),\\
\hatt&=\max\{t\in[M^d]:\imath_{q,f(q)}(x^\tau(\Gamma^{-1}(t));y^\tau)\geq \lambda\}.
\end{align*}
\end{algorithmic}
\end{algorithm}

Our non-asymptotic bound states as follows.
\begin{theorem}
\label{fbl:ach:adaptive}
Given any $(d,M)\in\bbR_+\times\bbN$, for any $q\in[0,1]$ and $\lambda\in\bbR_+$, Algorithm \ref{procedure:adapt} is an $(l,d,\frac{1}{M},\varepsilon)$-adaptive query procedure where
\begin{align}
l&\leq \mathbb{E}[\tau_1(\bX^\infty,Y^\infty)],\\
\varepsilon&\leq(M^d-1)\Pr\{\tau_1(\bX^\infty,Y^\infty)\geq \tau_2(\bX^\infty,Y^\infty)\},
\end{align}
where the expectation and probability are calculated with respect to the probability distribution $P_{\bX^\infty,Y^\infty}^{\calA^\infty,\Gamma^{-1}(1)}$ and $\calA^n$ refers to the queries in Algorithm \ref{procedure:adapt}.
\end{theorem}
In the proof of Theorem \ref{fbl:ach:adaptive}, we use the proof techniques on variable length feedback code in \cite{polyanskiy2011feedback} to analyze the performance of the query procedure in Algorithm \ref{procedure:adapt}. 

Note that although the result in Theorem \ref{fbl:ach:adaptive} holds for any $M$ and any discrete memoryless channel, it is hard to compute exactly in general. To obtain insights on the fundamental limit of adaptive querying, we derive second-order approximation to the non-asymptotic performance of the termination version of Algorithm \ref{procedure:adapt}, as outlined in Algorithm \ref{procedure:adapt2}. By termination, here we simply mean that the adaptive query procedure is not run with a certain probability (cf.~\cite{polyanskiy2011feedback}). This way, we can balance the complexity (i.e., the stopping time) and the performance (i.e., the achievable resolution), especially at relatively large excess-resolution probabilities.

\begin{algorithm}[bt]
\caption{Adaptive query procedure with Termination}
\label{procedure:adapt2}
\begin{algorithmic}
\REQUIRE Four parameters $(M,p,\lambda,\varepsilon)\in\bbN\times(0,1)\times\bbR_+\times(0,1)$
\ENSURE An estimate $(\hats_1,\ldots,\hats_d)\in[0,1]^d$ of a $d$-dimensional target variable $(s_1,\ldots,s_d)\in[0,1]^d$\\
\hrulefill
\STATE Generate a random number $Z$ according to the Bernoulli distribution where $Z$ takes value $1$ with probability $\varepsilon$
\IF {$Z=1$}
\STATE End the algorithm and output $\hats_i=0.5$ for all $i\in[d]$
\ELSE
\STATE Run Algorithm 2 with parameters $(M,p,\lambda)$ 
\ENDIF 
\end{algorithmic}
\end{algorithm}

Given Lipschitz continuous function $f$ and measurement-dependent channels $\{P_{Y|X}^q\}_{q\in[0,1]}$, the channel ``capacity" is defined as
\begin{align}
C_f&:=\max_{q\in[0,1]} \mathbb{E}[\imath_{q,f(q)}(X;Y)]\label{def:capacity},
\end{align}
where $(X,Y)\sim \mathrm{Bern}(q)\times P_{Y|X}^{f(q)}$.
The asymptotic approximation to the performance of Algorithm \ref{procedure:adapt2} is stated as follows.
\begin{theorem}
\label{second:fbl:adaptive}
For any $(l,d,\varepsilon)\in\bbR_+\times\bbN\times[0,1)$,
\begin{align}
-\log\delta^*_\rma(l,d,\varepsilon)\geq \frac{lC_f}{d(1-\varepsilon)}+O(\log l)\label{md:adaptive}.
\end{align}
\end{theorem}
The proof of Theorem \ref{second:fbl:adaptive} uses Theorem \ref{fbl:ach:adaptive} similarly to \cite{polyanskiy2011feedback} and uses the change-of-measure technique to replace a measurement-dependent channel with a measurement-dependent channel. In particular, the $\varepsilon=0$ case corresponds to the asymptotic performance of Algorithm \ref{procedure:adapt}.

We make two additional remarks. Firstly, a converse bound is necessary to establish the optimality of any adaptive query procedure under a measurement-dependent channel. However, a converse is elusive, since as pointed out in \cite{kaspi2018searching}, under the measurement-dependent channel, each noisy response $Y_i$ depends not only on the target vector $\bS$, but also the previous queries $\calA^{i-1}$ and noisy responses $Y^{i-1}$. This strong dependency makes it difficult to directly relate the current problem to channel coding with feedback~\cite{horstein1963sequential}. New ideas and techniques are required to obtain a converse result.

Another remark concerns a measurement-independent channel, i.e., the Lipschitz continuous function satisfies that $f(q)=\alpha$ with $\alpha\in[0,1]$ for all $q\in[0,1]$. In this case, the adaptive 20 questions problem is closely related to channel coding with feedback~\cite{jedynak2012twenty}. Therefore, using the non-asymptotic results on channel coding with feedback in \cite{polyanskiy2011feedback}, we conclude that the minimal achievable resolution $\delta^*_{\rma,\rm{mi}}(l,d,\varepsilon)$ satisfies
\begin{align}
-\log\delta^*_{\rma,\rm{mi}}(l,d,\varepsilon)=\frac{lC_{\rm{mi}}}{d(1-\varepsilon)}+O(\log l),\label{mi:adaptive}
\end{align}
where $C_{\rm{mi}}:=\max_{q\in[0,1]}\mathbb{E}[\imath_{q,\alpha}(X;Y)]$ is the capacity of the above measurement-independent channel.

\subsection{Comparisons with other Adaptive Query Procedures over a Measurement-Dependent BSC}
In the following, we specialize Theorem \ref{second:fbl:adaptive} to a measurement-dependent BSC with parameter $\nu$ and the Lipschitz continuous function $f(\cdot)$ such that $\max_{q\in[0,1]}\nu f(q)\leq 0.5$. Given any $q\in[0,1]$, let $\beta(\nu,q):=q(1-\nu f(q))+(1-q)\nu f(q)$. The capacity of the measurement-dependent BSC with parameter $\nu$ under the Lipschitz continuous function $f$ is given by
\begin{align}
C_f(\nu)
&=\max_{q\in[0,1]}\Big(h_\rmb(\beta(\nu,q))-h_\rmb(\nu f(q))\Big),
\end{align}
where $h_\rmb(p)=-p\log(p)-(1-p)\log(1-p)$ is the binary entropy function. Furthermore, let
\begin{align}
C_f^{\mathrm{sortPM}}(\nu)=\log (2)-h_\rmb(\nu f(0)).
\end{align}

Theorem \ref{second:fbl:adaptive} implies that for a measurement-dependent BSC with parameter $\nu$, Algorithm \ref{procedure:adapt2} achieves the asymptotic resolution decay rate
$\frac{lC_f(\nu)}{d(1-\varepsilon)}$ for a $d$-dimensional target when tolerating an excess-resolution probability of at most $\varepsilon$. In contrast, for $d=1$, the sorted posterior matching (PM) algorithm in \cite{chiu2016sequential} achieves the resolution decay rate $C_f^{\mathrm{sortPM}}(\nu)$ regardless of the excess-resolution probability $\varepsilon$. 

Note that $C_f^{\mathrm{sortPM}}(\nu)=C_f(\nu)$ if the Lipschitz continuous function $f(\cdot)$ is a constant value function. In other words, for a measurement-independent channel, unless for vanishing excess-resolution probability, i.e., $\varepsilon=0$, Algorithm \ref{procedure:adapt2} has a strict larger asymptotic resolution decay rate than the sorted PM algorithm due to the termination strategy which helps reduce the average stopping time and keep the same asymptotic resolution decay rate. Furthermore, Algorithm \ref{procedure:adapt2} has a lower time complexity $O(\frac{1}{\delta})$ compared with $O(\frac{-\log\delta}{\delta})$ of the sorted PM for a target resolution $\delta$ which usually tends to $0$ quickly with respect to the number of queries.

\begin{figure}[tb]
\centering
\includegraphics[width=.8\columnwidth]{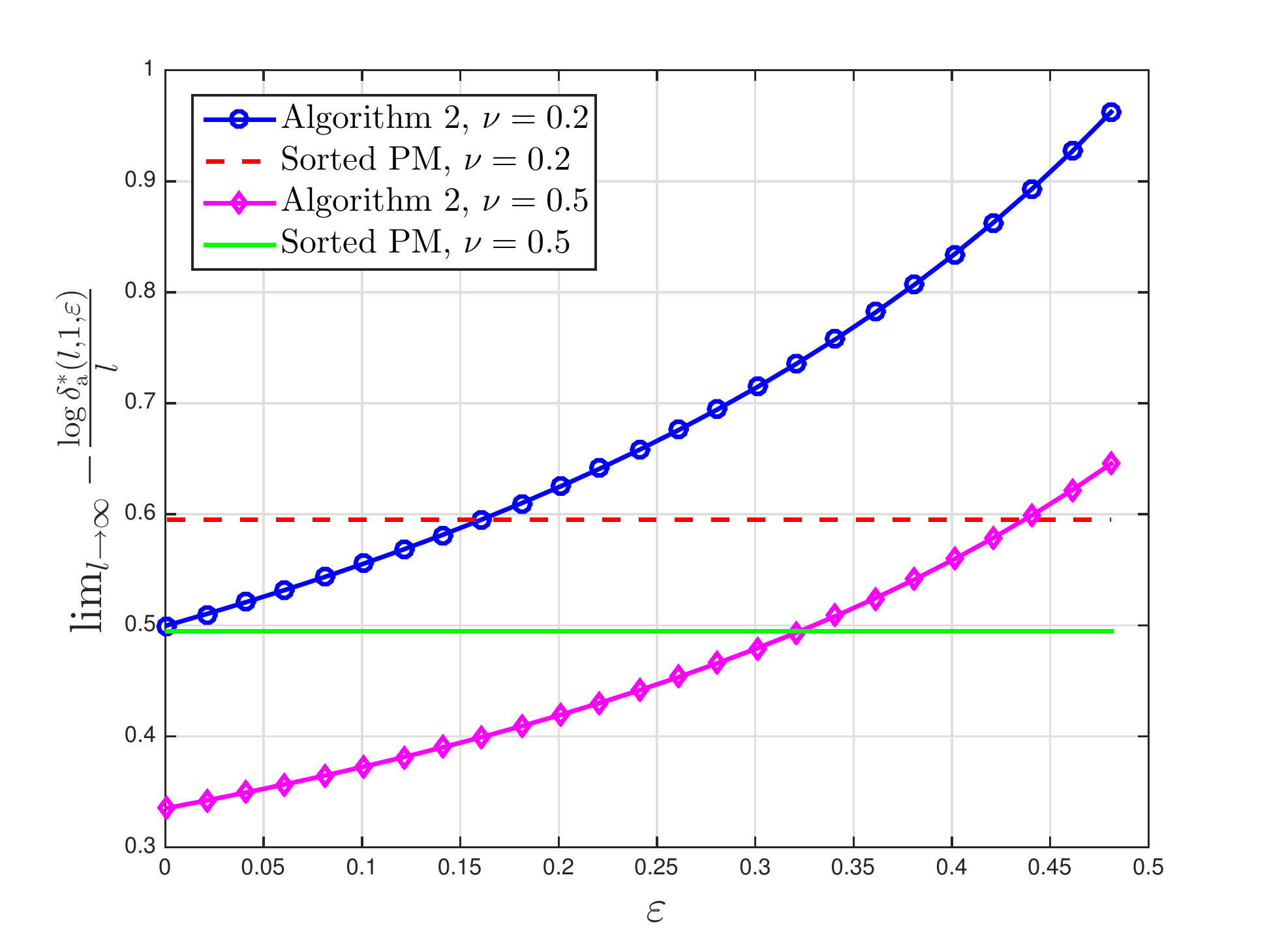}
\caption{Comparison of the asymptotic resolution decay rate of Algorithm \ref{procedure:adapt2} and the sorted PM algorithm in \cite{chiu2016sequential} for a measurement-dependent BSC with parameter $\nu$ over a range of tolerable excess-resolution probability $\varepsilon$. We consider the Lipschitz continuous function $f(q)=0.3q+0.1$.}
\label{com_adap_asymp}
\end{figure}

For a measurement-dependent channel, the Lipschitz continuous function $f(q)$ is \emph{not} a constant value function. If $f(q)$ decreases in $q$, we have $C_f(\nu)\geq C_f^{\mathrm{sortPM}}(\nu)$. Therefore, Algorithm \ref{procedure:adapt2} usually outperforms the sorted PM algorithm in this case. A measurement-dependent channel with $f(q)$ decreasing in $q$ corresponds to the scenario where the oracle penalizes a query more if the query inspects a smaller region and is thus closer to the true answer, which is motivated by practical searching problems where fine-grained search can be more error-prone. On the other hand, if $f(q)$ increases in $q$, we have $\frac{lC_f(\nu)}{d(1-\varepsilon)}\geq C_f^{\mathrm{sortPM}}(\nu)$ for large values of $\varepsilon$ and small values of $\nu$ if $f(q)$ increases in its parameter $q$.  Therefore, the asymptotic resolution decay rate of Algorithm \ref{procedure:adapt2} can also be larger than the sorted PM algorithm for a measurement-dependent channel with $f(q)$ increasing in $q$. See Figure \ref{com_adap_asymp} for a numerical comparison of the asymptotic resolution decay rates for the latter case.
%of Algorithm \ref{procedure:adapt2} and the sorted PM algorithm for the latter case.

One can apply the termination strategy in Algorithm \ref{procedure:adapt2} to any other adaptive query procedure with vanishing excess-resolution probability including the sorted PM algorithm to improve the asymptotic resolution decay rate. In particular, the sorted PM algorithm with termination achieves the asymptotic resolution decay rate $\frac{C_f^{\mathrm{sortPM}}(\nu)}{1-\varepsilon}$ for any termination probability $\varepsilon\in(0,1)$. Thus, with termination, the sorted PM can achieve a larger asymptotic resolution decay rate than Algorithm \ref{procedure:adapt2} for a measurement-dependent BSC if $C_f^{\mathrm{sortPM}}(\nu)\geq C_f(\nu)$, which holds true if $f(q)$ increases in $q$. 

With the comparisons above, we find that Algorithm \ref{procedure:adapt2} and the sorted PM algorithm can outperform each other in certain cases. Therefore, a deeper investigation on the fundamental limit of adaptive querying is required to uncover the non-asymptotic and asymptotic performance of an optimal adaptive query procedure, even just for a measurement-dependent BSC.

\subsection{Numerical Illustration}
\begin{figure}[tb]
\centering
\begin{tabular}{c}
\includegraphics[width=.82\columnwidth]{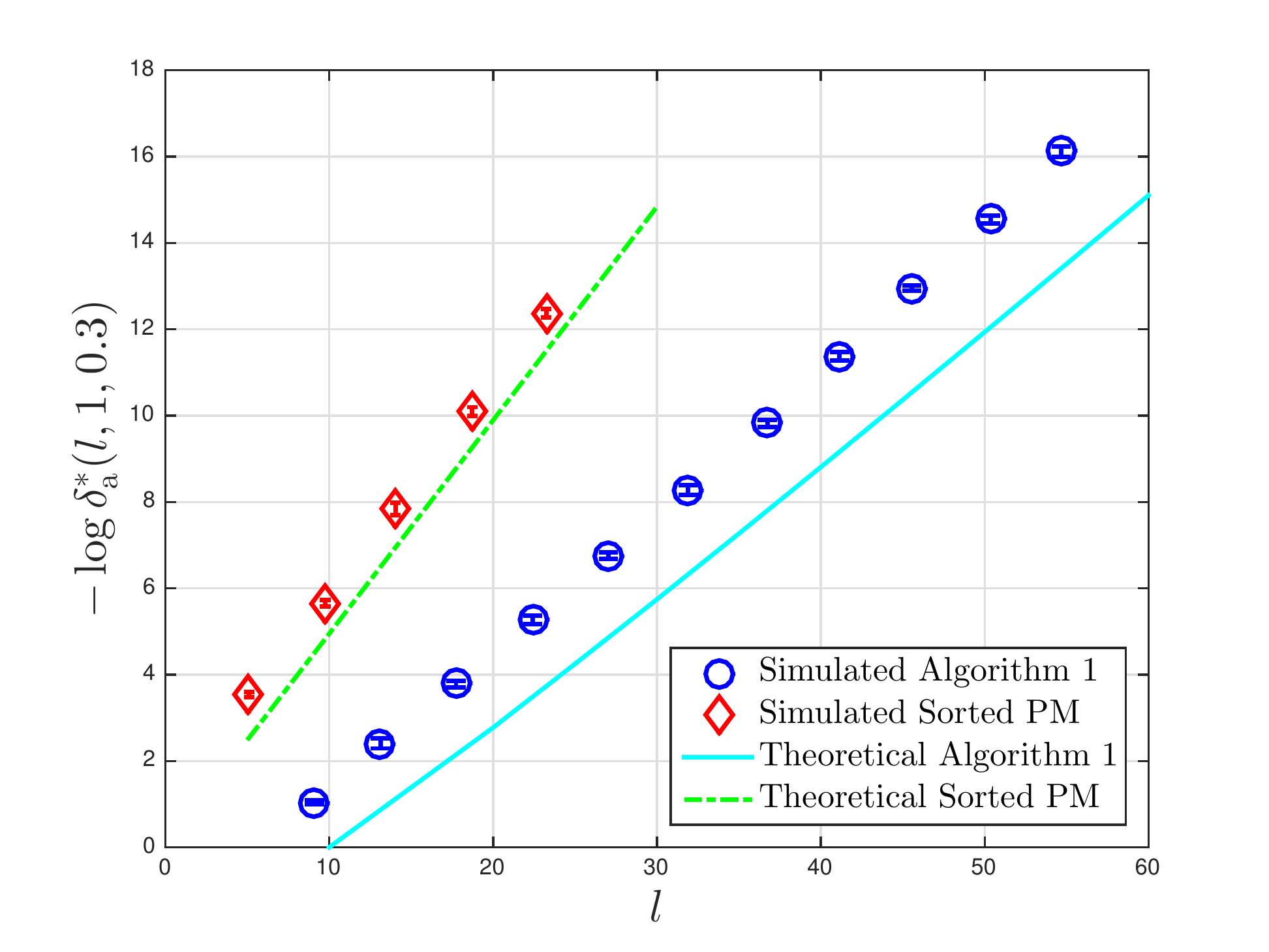}\\
{(a) $f(q)=0.1+0.3q$}\\
\includegraphics[width=.82\columnwidth]{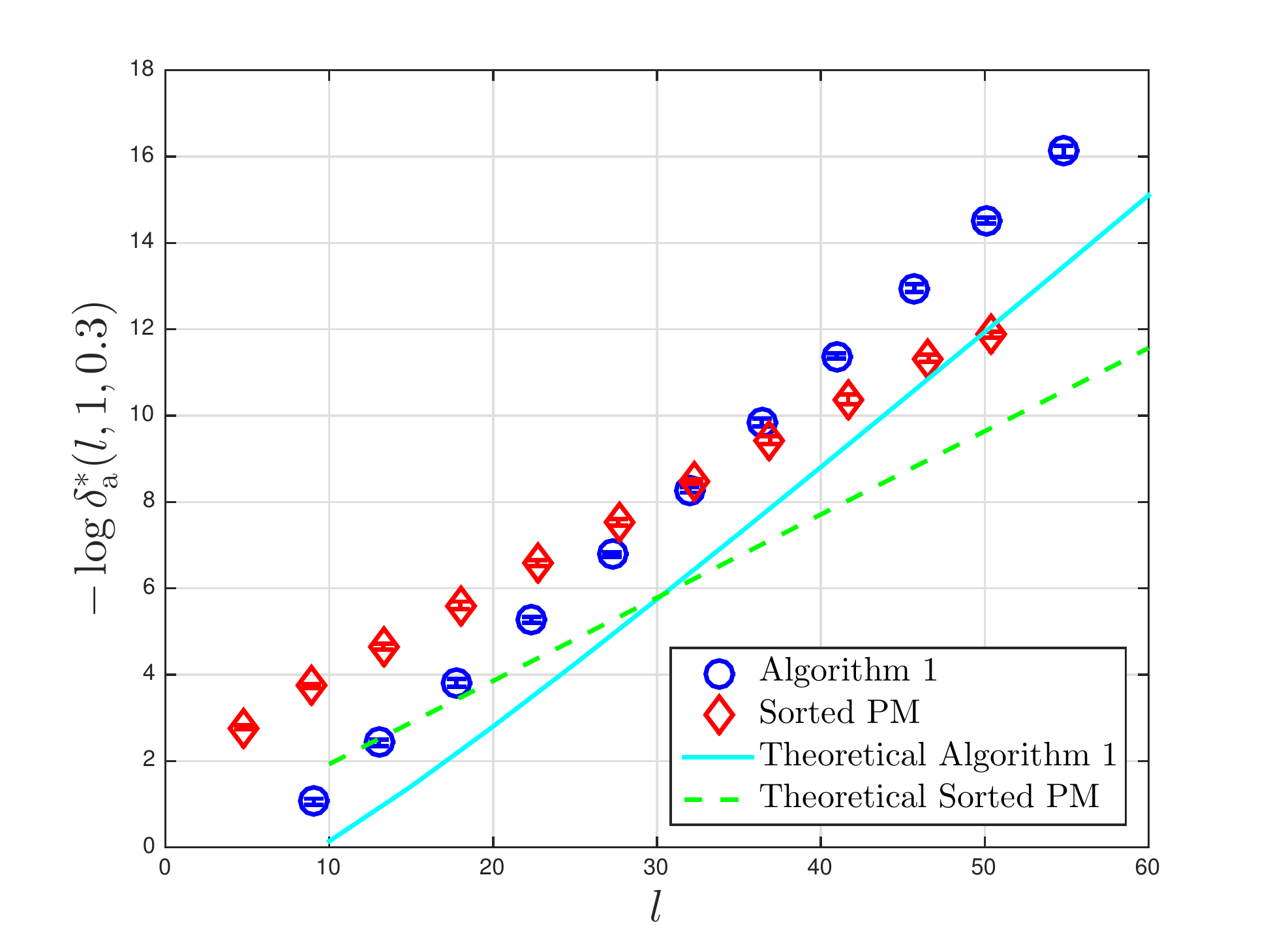}\\
 {(b) $f(q)=0.4-0.3q$}
\end{tabular}
\caption{Achievable resolution of adaptive query procedures over a measurement-dependent BSC with parameter $\nu=0.5$ and different Lipschitz continuous functions. The symbols correspond to Monte Carlo simulations of adaptive query procedures and the solid lines correspond to theoretical lower bounds. The error bars for the simulated results denote three standard deviations about the mean. 
}
\label{sim_adap}
\end{figure}

We numerically illustrate the achievable resolutions of Algorithm \ref{procedure:adapt} and the sorted PM algorithm for a one-dimensional uniformly distributed target variable $S$ over $[0,1]$. The simulation settings and results are provided in Figure \ref{sim_adap}, which validate our theoretical analyses. Note that we do not simulate the termination version in Algorithm \ref{procedure:adapt2} because i) we would like the comparison to be fair because the sorted PM algorithm does not use termination and ii) to demonstrate the benefit of termination of Algorithm \ref{procedure:adapt2}, we need to run simulations with very large number of queries, which is computationally intractable.

\section{Conclusion}
We studied the achievable resolution of adaptive query procedures for a $d$-dimensional target using the framework of 20 questions estimation under the measurement-dependent noise channel. Our results implied the benefit of adaptivity for any measurement-dependent discrete memoryless channel. In future, one can nail down the exact fundamental limit of adaptive query procedures to either prove the tightness of our result under certain conditions or show that our bound can be further improved in general. It is also interesting to generalize the algorithm and the analysis of the sorted PM algorithm to other channels beyond the BSC. Finally, one can propose low-complexity (preferably polynomial in the number of queries) query procedures that achieve our derived theoretical results.

\newpage
\bibliographystyle{IEEEtran}
\bibliography{IEEEfull_lin}

\end{document}